\documentclass[submit,ams]{jsme-tj-movic}
\usepackage{wrapfig}
\usepackage{newtxtext,newtxmath}
\usepackage{amsmath,bm}
\usepackage[authoryear,round,sort&compress]{natbib}
\usepackage{mathtools}
\def\newblock{\relax}

\makeatletter
\newcommand{\gv}{\bm}

\newcommand{\param}{{\gv q}}
\newcommand{\oparam}{{\gv q^{\ast}}}
\newcommand{\sss}{{\gv s}}
\newcommand{\ee}{{\gv e}}

\newcommand{\ff}{{\gv f}}
\newcommand{\xx}{{\gv x}}
\newcommand{\BB}{{\gv B}}
\newcommand{\JJ}{{\gv J}}
\newcommand{\SSS}{{\mathbb S}}
\newcommand{\clD}{{\mathcal D}}
\newcommand{\pdata}{{\bar p}}
\newcommand{\FPE}{\mathop{\text{\sf FPE}}}

\renewcommand{\div}{\nabla}

\newcommand{\setdef}[2]{\{#1\;|\;#2\}}

\def\@pd[#1]#2#3{\frac{\partial^{#1} #2}{\partial {#3}^{#1}}}
\def\pd{\@ifnextchar[{\@pd}{\@pd[]}}
\def\@od[#1]#2#3{\frac{d^{#1} #2}{d {#3}^{#1}}}
\def\od{\@ifnextchar[{\@od}{\@od[]}}

\def\@npd[#1]#2#3{\frac{\Delta^{#1} #2}{\Delta {#3}^{#1}}}
\def\npd{\@ifnextchar[{\@npd}{\@npd[]}}

\newcommand{\leqn}[1]{\label{eqn:#1}}
\newcommand{\reqn}[1]{(\ref{eqn:#1})}
\newcommand{\lfig}[1]{\label{fig:#1}}
\newcommand{\rfig}[1]{Fig.\ \ref{fig:#1}}

\newcommand{\lsec}[1]{\label{sec:#1}}
\newcommand{\rsec}[1]{Section\ \ref{sec:#1}}

\let\textscript=\relax
\makeatother

\MoViCtitle{MoViC2020}%
{The 15th International Conference on Motion and Vibration}%
{\copyright\ 2020 The Japan Society of Mechanical Engineers}

\title{Parameter Estimation via Fokker--Planck Type Residual:
Application to Linear Stationary Random Vibration}

\vol{1}\no{1}\field{XXX}\received{2014/3/3}{XX-XXXX}

\authors{\underline{Katsutoshi YOSHIDA$^1$$^\ast$}, Yoshikazu YAMANAKA$^1$}
\affiliates{%
  $^1$Department of Mechanical Systems Engineering,
  Utsunomiya University, 7-1-2 Yoto, Utsunomiya,
  Tochigi 321-8585, Japan}

\emails{%
  $^\ast$yoshidak@cc.utsunomiya-u.ac.jp\\
  yyamanaka@cc.utsunomiya-u.ac.jp}%

\begin{document}

\maketitle

\begin{abstract}
In this study, we propose a new method that is useful for estimating unknown parameter values of stochastic differential equation (SDE) models, based on probability density function (PDF) data measured from random dynamical systems. As our method does not require explicit description of PDF, it can be applied to the SDE models even when their PDFs are hardly derived in explicit forms due to multiplicative-noise terms, nonlinear terms, and so on. Therefore,
our method is expected to provide a versatile tool to dynamically parameterize measured PDF data. In our proposed method, it is assumed that a measured PDF is obtained from a random dynamical system whose structure is described by a known SDE model with unknown parameter values. With the help of It\^o calculus, the Fokker--Planck equation (FPE) is derived from the SDE model. The measured PDF and a candidate of parameter values are substituted into the FPE to calculate a FPE residual. Our method is applied to two random vibration systems. Their FPE residuals tend to zero as the parameter values tend to exact values, showing that our proposed FPE residual can be utilized for unknown parameter estimation of SDE models.
\end{abstract}

\begin{keywords}
Parameter, Estimation, Fokker--Planck equation, Probability density function
\end{keywords}

\section{Introduction}

\begin{wrapfigure}{r}{.42\hsize}
  \captionwidth\hsize
  \centering
  \includegraphics[width=.95\hsize]{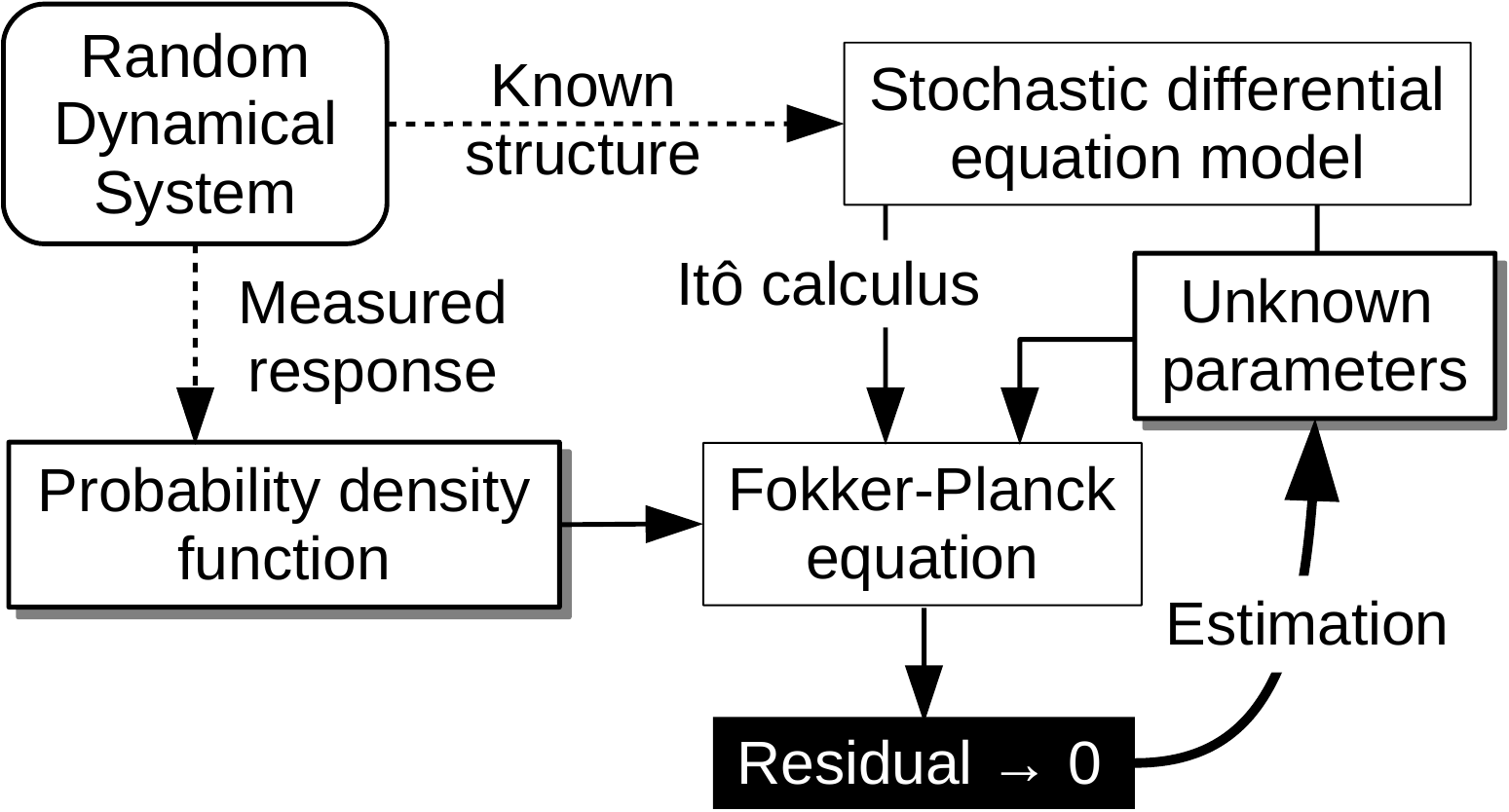}
  \caption{Conceptual diagram of our proposed method.}
  \lfig{concept}
\end{wrapfigure}

Efficient and accurate modeling of random dynamical systems is an old and new issue in the fields of
physics \citep{HATJISPYROS200771},
engineering \citep{POULIMENOS2006763,Yoshida_2019bicycle},
economics \citep{Bhattacharya2003},
and so on.

In this study, we propose a new method that estimates unknown parameter values of SDE models based on measured PDF data of random dynamical systems.
As shown in \rfig{concept}, our proposed method assumes that a measured PDF is obtained from a random dynamical system whose structure is described by a known SDE model with unknown parameter values. The SDE model is transformed into the corresponding FPE with the help of It\^o calculus. The measured PDF and a candidate of parameter values are substituted into the FPE to calculate a FPE residual. The resulting FPE residual is expected to tend to zero as the parameter values tend to exact values.

Our continuous time domain approach clearly differs from other typical previous methods using discrete time models and/or frequency domain techniques \citep{HATJISPYROS200771,POULIMENOS2006763}.
Moreover, our method has the advantage that it does not require explicit description of PDF; it can be applied to the SDE models whose PDFs are hardly derived in explicit forms due to multiplicative-noise terms, nonlinear terms, and so on.

In this study, our proposed method is tested on two linear random vibration systems. The resulting FPE residuals tend to zero as the parameter values tend to exact values, meaning that our proposed method is capable of estimating unknown parameters of SDE models.

The rest of the paper is organized as follows:
Section 2 describes our problem formulation.
Section 3 describes our proposed method of parameter estimation.
Section 4 demonstrates capability of our proposed method by numerical examples.
Section 5 concludes our study.

\section{Problem formulation}

We consider an $n$-dimensional time-invariant random dynamical system described by a Stratonovich SDE as
\begin{equation}
  d\xx = \ff^{\circ}(\xx;\param)dt + G(\xx;\param)\circ d\BB
  ,\quad
  \xx(0)=\xx_0\in R^n
  ,
  \leqn{SSDE}
\end{equation}
where $t$ is time, $\xx$ is the random state vector, $\ff^{\circ}$ is an $n$-dimensional vector-valued function, $G(\xx)$ is an $n\times m$ matrix-valued function, $\BB$ is an $m$-dimensional standard Brownian motion, and $\param$ is a $Q$-dimensional parameter vector.

In this study, we impose the following assumptions on our problem:
\begin{itemize}
  \item The system is stable.
  \item The system functions $\ff^{\circ}$ and $G$ are structurally known.
  \item The value of the parameter vector $\param$ is unknown.
  \item A stationary PDF data $\pdata(\xx)$ of the state $\xx$ is obtained for a certain $\param$.
\end{itemize}
Our problem is to estimate $\param$ that makes the SDE \reqn{SSDE} reproduce the measured $\pdata(\xx)$ shape in the sense of least squares.

\section{Method of parameter estimation}

\lsec{method}

\subsection{Fitting measure based on FPE residual}

Although the Stratonovich SDE \reqn{SSDE} provides a proper stochastic description of classical mechanics with random fluctuations, it is incompatible with It\^o calculus \citep{gardiner2004handbook} that we use in this study.
To make it compatible with It\^o calculus, we transform \reqn{SSDE} to the It\^o SDE,
\begin{equation}
  d\xx = \ff(\xx;\param)dt+G(\xx;\param)d\BB
  ,\quad
  \xx(0)=\xx_0 \in R^n,
  \leqn{ISDE}
\end{equation}
which is mathematically equivalent to \reqn{SSDE} through
\begin{equation}
  f_i(\xx;\param) =
  f^{\circ}_i(\xx;\param)
  + \frac12 \sum_{k=1}^{n}\sum_{j=1}^{m}
  G_{kj}(\xx;\param)\pd{G_{ij}}{x_k}(\xx;\param)
  ,\quad i=1,\cdots,n
  .
\end{equation}
Hereafter, $(\cdot)_i$ and $(\cdot)_{ij}$ denote $i$th and $(i,j)$th components of vector and matrix, respectively.

According to It\^o calculus \citep{gardiner2004handbook}, the PDF of the solution of \reqn{ISDE} satisfies the following FPE:
\begin{equation}
  \FPE(p(\xx,t);\param):=
  \pd{p(\xx,t)}{t} + \div \JJ(p(\xx,t);\param) = 0,\quad
  p(\xx,0)=\delta(\xx-\xx_0)
  :=\prod_{i=1}^{n} \delta\big(x_i-(\xx_0)_i\big),
  \leqn{FPE}
\end{equation}
where $\delta(\xx)$ is the $n$-dimensional Dirac's delta function and
\begin{equation}
  \div \JJ(p(\xx);\param) := \sum_{i=1}^{n}\pd{J_i(p(\xx);\param)}{x_i}
\end{equation}
is the divergence of the vector field $\JJ(p(\xx);\param)$ of $\xx$ (called a probability current) whose components are given by
\begin{equation}
  J_i(p(\xx);\param) :=
  f_i(\xx;\param)p(\xx)
  -\frac12\sum_{j=1}^{n}\pd{}{x_j}\Big\{(G(\xx;\param)G(\xx;\param)^T)_{ij}p(\xx)\Big\}
  ,\quad i=1,\cdots,n
  .
\end{equation}

Letting $p(\xx,t;\param)$ be the exact solution of \reqn{FPE} and $\pdata(\xx,t)$ be an estimated solution, we propose a {\it PDF-fitness} as
\begin{equation}
  E(\pdata(\xx,t);\param)
  := \int_{\clD}\left\{
    \FPE(\pdata(\xx,t);\param)
  \right\}^2 d\xx,
  \leqn{cost}
\end{equation}
where $\clD$ is an appropriate state-space domain and $\FPE(\pdata(\xx,t);\param)$ is a FPE residual.
Obviously, $E(\pdata(\xx,t);\param)=0$ for $\pdata(\xx,t)=p(\xx,t;\param)$.
Furthermore, in this study, we expect the following asymptotic property:
\begin{equation}
  E(\pdata(\xx,t);\param)\to 0\quad(\pdata(\xx,t)\to p(\xx,t;\param)).
\end{equation}

Based on the above, we can solve the optimization problem:
\begin{equation}
  \param^\ast = \arg\min_{\param} E(\pdata(\xx,t);\param),
  \leqn{opt}
\end{equation}
where $\pdata(\xx,t)$ is substituted by the measured PDF data in the following sections.

In addition, since this study assumes that the system is time-invariant and stable, the FEP has a stationary solution, i.e., $p(\xx,t)\to p(\xx)$ ($t\to\infty$); this yields $\partial p/\partial t=0$ and simplifies the FPE residual to the stationary version:
\begin{equation}
  \FPE(\pdata(\xx);\param):=\div \JJ(\pdata(\xx);\param).
  \leqn{stationaryFPE}
\end{equation}
In the following sections, we use the stationary \reqn{stationaryFPE} to solve the optimization problem \reqn{opt} and estimate unknown $\param$ with the stationary PDF data.

\subsection{Numerical implementation}
\lsec{implementation}

We consider a time-series of the state vector $\xx$ of length $N$ measured from the random dynamical system \reqn{SSDE} as
\begin{equation}
  \xx(k) := \xx(t_k)
  ,\quad
  t_k := t_0 + k\Delta t
  ,\quad
  k = 0,\cdots,N-1
\end{equation}
where $\Delta t$ is a sampling interval.
A sufficiently large initial time $t_0$ is taken to obtain a stationary response.

\subsubsection{Quantization of state space}

A hyperrectanglar state-space domain $\clD$ is taken as
\begin{equation}
  \clD := \prod_{i=1}^{n}[\underline x_i,\overline x_i) \subset R^n
  ,
\end{equation}
where $\prod$ denotes Cartesian product of sets and $x\in[\underline x_i,\overline x_i)$ denotes $\underline x_i \leq x < \overline x_i$.
The domain $\clD$ is divided into a direct sum of $S^n$ uniform hyperrectangular cells of the form:
\begin{equation}
  \clD[\sss] := \prod_{i=1}^{n}
  \left[
    x_i[s_i]-\frac{\Delta_i}{2}
    ,
    x_i[s_i]+\frac{\Delta_i}{2}
  \right)
  ,\quad
  \Delta_i := \frac{\overline x_i - \underline x_i}{S}
  ,
\end{equation}
where $s_i$ is the $i$th component of the $n$-dimensional index:
\begin{equation}
  \sss \in \SSS
  :=
  \setdef{(s_1,\cdots,s_n)}{1\leq s_i \leq S\;\text{for}\; 1\leq \forall i\leq n}
\end{equation}
and
\begin{equation}
  x_i[s_i] := \underline x_i + \left(s_i-\frac12\right)\Delta_i
\end{equation}
is the $i$th component of the center point $\xx[\sss]$ of the $\sss$th cell.

In this gridded state-space domain, we designate the value of a function $F(\xx)$ at a grid-point $\xx[\sss]$ as
\begin{equation}
  F[\sss] := F(\xx[\sss])
\end{equation}
and approximate its partial derivative with a numerical derivative in the following form:
\begin{equation}
  \pd{F}{x_i}(\xx[\sss]) \approx
  \npd{F}{x_i}[\sss]
  :=
  \begin{dcases}
    \{F[\sss+\ee_i]-F[\sss]\}(\Delta_i)^{-1}
    & (s_i=1),
    \\
    \{F[\sss]-F[\sss-\ee_i]\}(\Delta_i)^{-1}
    & (s_i=S),
    \\
    \{F[\sss+\ee_i]-F[\sss-\ee_i]\}(2\Delta_n)^{-1}
    & (\text{Otherwise}),
  \end{dcases}
\end{equation}
where $\ee_i$ denotes the $i$th standard basis vector in $\SSS$.

\subsubsection{Construction of PDF data}
\lsec{pdfdata}

We count the frequency $N[\sss]$ of $\xx(k)\in\clD[\sss]$ for each $\sss$ and obtain a numerical probability density $\pdata[\sss]$ at each grid-point $\xx[\sss]$ as follows.
\begin{equation}
  \pdata[\sss] := \frac{N[\sss]}{N\Delta_1\cdots\Delta_n}
  ,\quad
  \sss\in\SSS,
\end{equation}
which we call a PDF data.

\subsubsection{Implementation of the PDF-fitness}

We numerically implement the FPE residual in \reqn{stationaryFPE} as
\begin{equation}
  \FPE[\pdata,\sss;\param]
  := \sum_{i=1}^{N}\npd{J_i[\pdata,\sss;\param]}{x_i}
\end{equation}
with a numerical probability current of the form:
\begin{equation}
  J_i[\pdata,\sss;\param] :=
    f_i[\sss;\param]\pdata[\sss]
    -\frac12\sum_{j=1}^{N}
      \npd{}{x_j} \left\{
        (G[\sss;\param]G[\sss;\param]^T)_{ij}\pdata[\sss]
      \right\}
  .
\end{equation}
Therefore, we have a numerical representation of the PDF-fitness in \reqn{cost} as
\begin{equation}
  E[\pdata;\param] := \sum_{\sss\in S}\{\FPE[\pdata,\sss;\param]\}^2
  \leqn{numcost}
\end{equation}
to solve our parameter estimation problem:
\begin{equation}
  \oparam = \arg\min_{\param}E[\pdata;\param],
  \leqn{numopt}
\end{equation}
based on the measured PDF $\pdata[\sss]$ with the known system structure \reqn{SSDE}.

\section{Application to random linear vibration systems}

In this final section, we demonstrate capability of our proposed method described in \rsec{method}.

\subsection{Random linear vibration system with additive noise}

To provide reliable tests on our method, we start with a linear vibration system whose PDF is explicitly known.
Specifically, we consider a linear vibration system with an additive noise only, given by
\begin{equation}
  \ddot x + c \dot x + k x = \nu\, w(t),
  \leqn{LVSa}
\end{equation}
where $c$ is a damping, $k$ is a stiffness, $\nu$ is a noise strength, and $w(t)$ is a standard Gaussian white noise.
Using a state vector $\xx := (x, \dot x)^T$ ($^T$ denotes transpose) and a parameter vector $\param:=(k,c,\nu)$, we rewrite \reqn{LVSa} to the SDE from with
\begin{equation}
  \ff^\circ(\xx;\param) :=
  \begin{bmatrix}
    0 & 1 \\
    -k & -c
  \end{bmatrix}
  ,\quad
  G(\xx;\param) :=
  \begin{bmatrix}
    0 \\
    \nu
  \end{bmatrix}
  ,\quad
  \BB:= B\;(\text{scalar})
  .
\end{equation}

We obtained a sample of time-series $\xx_n$ of length $N=10^7$ by numerically solving \reqn{LVSa} from $x(0)=\dot x(0)=0$ for $\param=\bar\param:=(1,0.5,1)$ with time step $\Delta t=0.01$.
To obtain a stationary data, the same length of initial response was skipped to store.
For numerical integration, a fourth-order Runge--Kutta--Gill method was used with the white noise term $w(t)$ simulated by $w(k) \approx W_k/\sqrt{\Delta t}$ ($k=0,1,\cdots$) where $W_k$ is normal pseudo-random numbers and $1/\sqrt{\Delta t}$ is the numerical factor required for integrating stochastic differential equations \citep{Higham.2012}.
\begin{figure}[t]
  \begin{minipage}[b]{.5\hsize}
    \centering
    \includegraphics[width=\hsize]{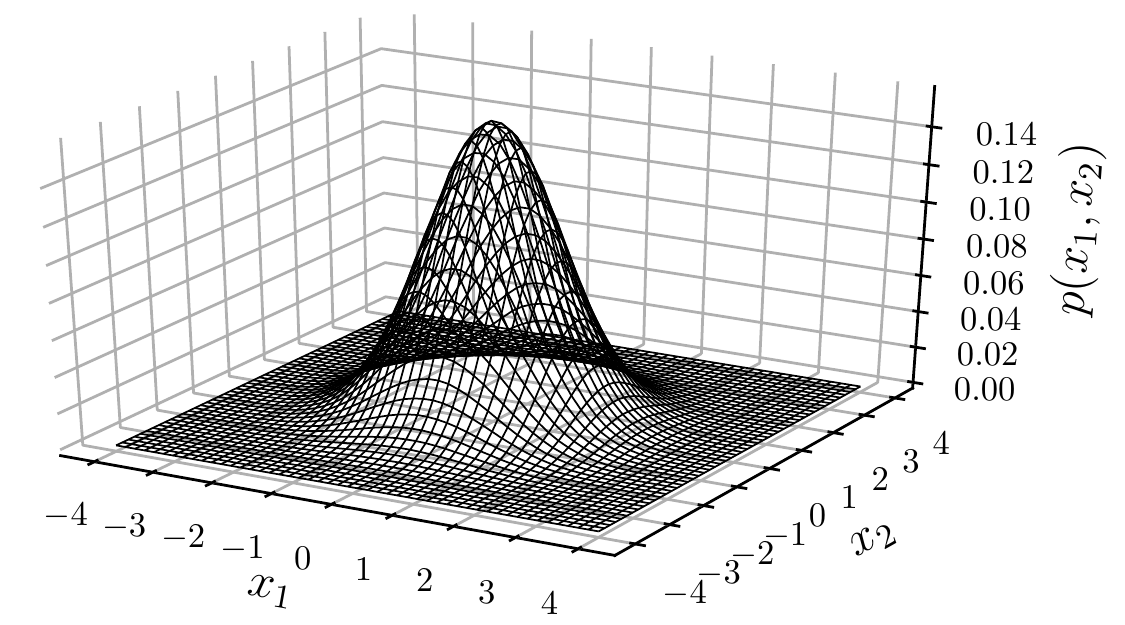}
    \caption{
      PDF data of linear vibration system \reqn{LVSa} on the $(x_1,x_2)$-domain $\clD=[-4,4)\times[-4,4)$ with the bin number $S=50$, which
      was numerically generated by means of the procedure described in \rsec{pdfdata}.
    }
    \lfig{PDF:LVSa}
  \end{minipage}%
  \begin{minipage}[b]{.5\hsize}
    \centering
    \includegraphics[width=\hsize]{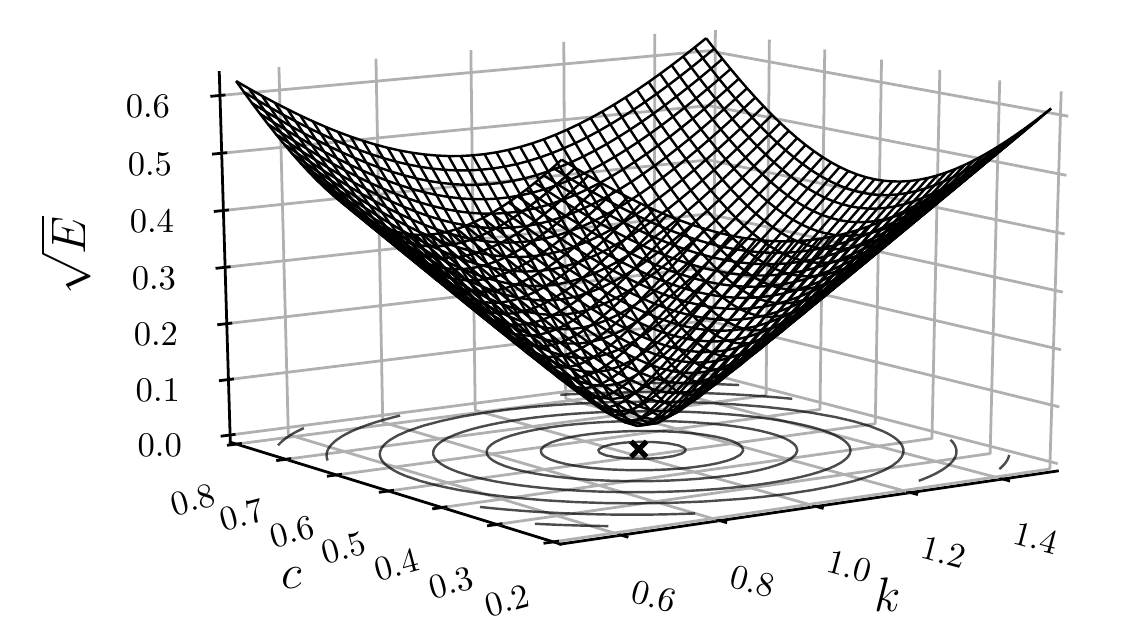}
    \caption{
      PDF-fitness value $E=E[\pdata;\param]$ with $\nu=\bar q_3=1$ fixed; the square root value is plotted.  On the plane $E=0$, the small cross mark indicates the exact parameter value $\bar\param$ and the curves indicate contours of $E$.
    }
    \lfig{cost3d_ck}
  \end{minipage}
\end{figure}
From this stationary $\xx_n$, the PDF data $\pdata[\sss]$ was numerically generated as shown in \rfig{PDF:LVSa}, by means of the procedure described in \rsec{pdfdata}, on the $(x_1,x_2)$-domain $\clD=[-4,4)\times[-4,4)$ with the bin number $S=50$.

\subsubsection{Estimation of independent parameters}

\rfig{cost3d_ck} plots the value of our proposed PDF-fitness $E=E[\pdata;\param]$ on $(k,c)$-parameter plane with $\nu=\bar q_3=1$ fixed; the square root value is plotted for ease of viewing.  On the plane $E=0$, the small cross mark indicates the exact parameter value $\bar\param$ and the curves indicate contours of $E$.
Obviously, our proposed $E$ forms a smooth and unimodal concave shape and its minimal point appears to be close to the exact $\bar\param$.
This implies that our optimization problem in \reqn{numcost} can be solved by gradient methods.
In this way, the unknown parameter values were estimated as $(k,c)=(0.99980749,0.49415439)$ with $E=4.529\times10^{-4}$; their estimation errors were $(-0.019\%,-1.169\%)$ of the exact $(k,c)=(1,0.5)$.

Therefore, the results show that our proposed method achieved a reasonable degree of accuracy of the unknown parameter estimation.

\subsubsection{Extraction of parameter dependency}

\begin{figure}[t]
  \begin{minipage}[b]{.5\hsize}
    \centering
    \includegraphics[width=\hsize]{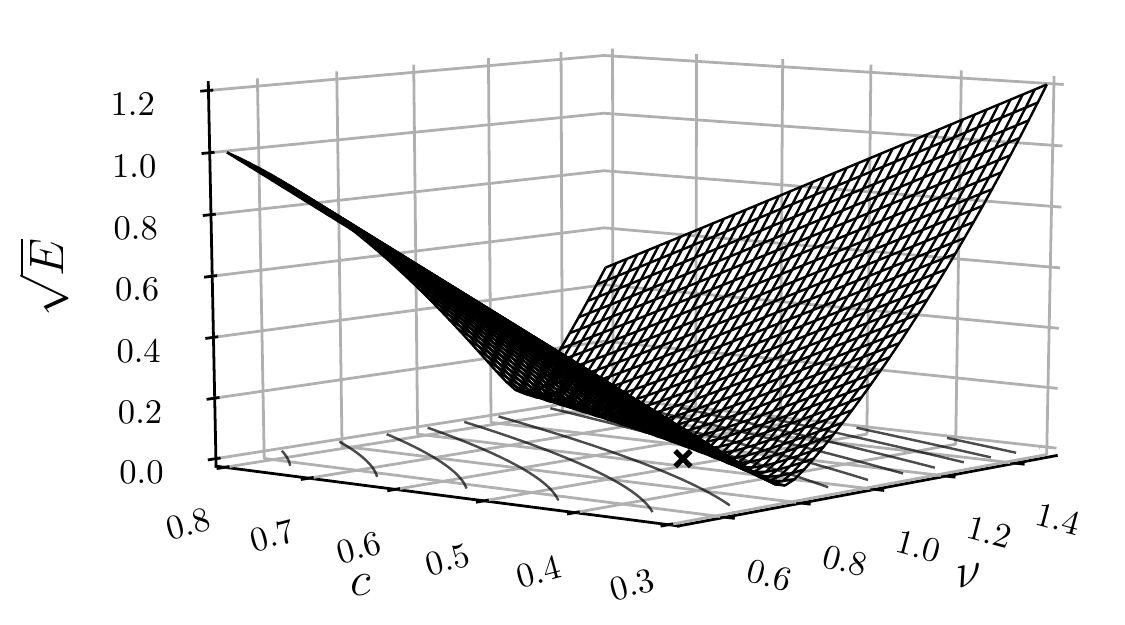}
    \caption{
      PDF fitness value $E=E[\pdata;\param]$ plotted on $(\nu,c)$-parameter plane with $k=\bar q_1=1$ fixed.
      The level set $c=\nu^2/2$ (or $\nu^2/(2c)=1$) appears as the bottom of the ravine shape of $E$.
    }
    \lfig{cost3d_nuc}
  \end{minipage}%
  \begin{minipage}[b]{.5\hsize}
    \centering
    \includegraphics[width=.92\hsize]{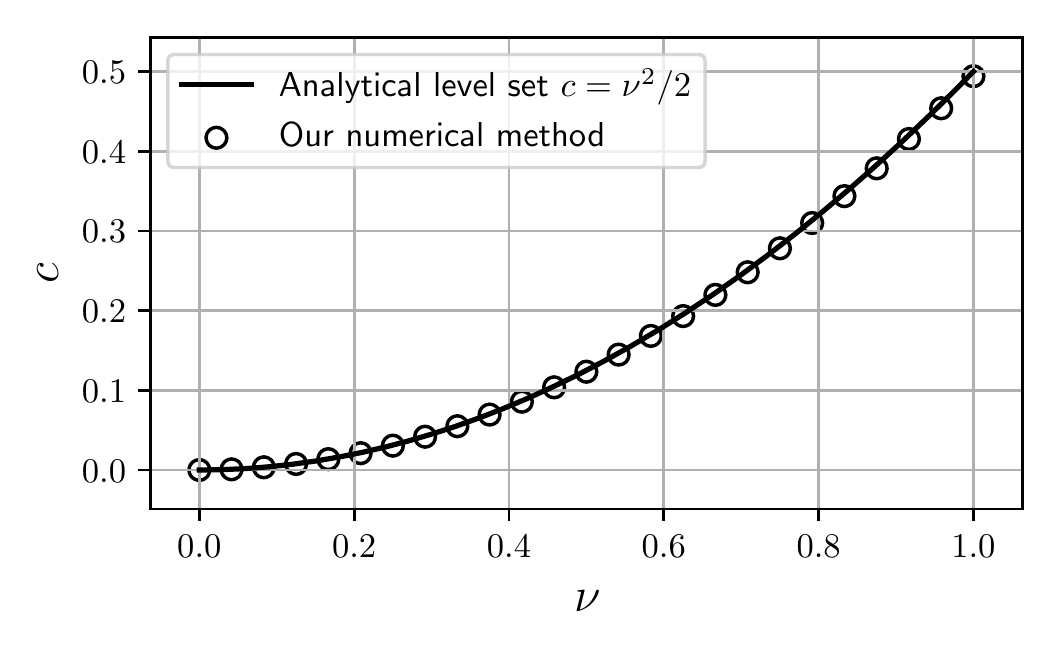}
    \caption{
      Numerically extracted level set from the PDF in \rfig{PDF:LVSa}.
      The small circles indicate our estimated $c$-value as a function of $\nu$ with $k=q_1=1$ fixed. The solid curve indicates the exact level set $c=\nu^2/2$.
    }
    \lfig{cost2d_nuc}
  \end{minipage}%
\end{figure}

Our method can also be applied to extracting parameter dependency.
As the system \reqn{LVSa} is linear and additive Gaussian, its stationary PDF becomes Gaussian and can be explicitly derived as \citep{Risken}
\begin{equation}
  p(\xx) =\frac{c\sqrt{k}}{\pi\nu^2}
  \exp\left\{-\frac{c}{\nu^2}(kx_1^2+x_2^2)\right\}.
  \leqn{LVSa:pdf}
\end{equation}
Such a Gaussian distribution is determined by only the following statistics:
\begin{equation}
  E[x_1]=E[x_2]=0, \quad
  V[x_1]=\frac{\nu^2}{2ck}, \quad
  V[x_2]=\frac{\nu^2}{2c}, \quad
  C[x_1,x_2]=0,
  \leqn{LVSa:stat}
\end{equation}
where $E[x]$ and $V[x]$ denote mean and variance of $x$, respectively, and $C[x,y]$ denotes covariance between $x$ and $y$.
In these statistics, only the variances depend on the system parameters and they have a common factor $\nu^2/(2c)$.
In other words, the shape of the PDF \reqn{LVSa:pdf} is parameterized by $\param=(k,c,\nu)$ under the constraint $\nu^2/(2c)=V[x_2]$. Particularly, in this example, since the PDF data was built for $\param=(1,0.5,1)$, our PDF-fitness $E$ must vanish along the level set $\nu^2/(2c)=V[x_2]=1^2/(2\times0.5)=1$.

\rfig{cost3d_nuc} shows the corresponding $E$-value on $(\nu,c)$-parameter plane with $k=\bar q_1=1$ fixed.
The abovementioned level set $c=\nu^2/2$ (or $\nu^2/(2c)=1$) appears as the bottom of the ravine shape of $E$. In \rfig{cost2d_nuc}, the small circles indicate our estimated $c$-value for given $k$ and $\nu$ values where $k$ is fixed to the exact $k=q_1=1$ and $\nu$ is taken at $25$ uniform grid points within the range $0\leq \nu\leq 1$. The solid curve indicates the exact level set $c=\nu^2/2$. In these results, the maximal estimation error of $c$ was less than $-2.81\%$ of the exact value.

In this way, our proposed method also achieved a reasonable degree of accuracy of extracting unknown parameter dependency.

\subsection{Random linear vibration system with both additive and multiplicative noises}

The second example is of the system whose PDF is not explicitly known. This is simply given by a linear vibration system subjected to both additive and multiplicative noises as follows.
\begin{equation}
  \ff^\circ(\xx;\param) :=
  \begin{bmatrix}
    0 & 1 \\
    -k & -c
  \end{bmatrix}
  ,\quad
  G(\xx;\param) :=
  \begin{bmatrix}
    0 & 0\\
    \nu_1 & \nu_2 x_1
  \end{bmatrix}
  ,\quad
  \BB:=
  \begin{bmatrix}
    B_1 \\
    B_2
  \end{bmatrix}
  \leqn{lin_a_m}
  ,
\end{equation}
where $B_1$ and $B_2$ are independent standard Brownian motions and $\nu_1$ and $\nu_2$ are the corresponding noise strengths. This system is linear but its explicit PDF is not yet known \citep{GITTERMAN2005309,ZORZANO1999109,Nakao.PhysRevE.58.1591}.

Using the same procedure and conditions as those in the first example, we generated the PDF data of \reqn{lin_a_m} for $\param=\bar\param:=(k,c,\nu_1,\nu_2)=(1,0.5,1,0.6)$.
Then, we applied our method to estimating $(\nu_1,\nu_2)$ values with $(k,c)=(\bar q_1,\bar q_2)$ fixed. The resulting estimation was  $(\nu_1,\nu_2)=(1.00762436,0.58836647)$ with the PDF-fitness $E=7.267\times10^{-4}$ and thus their estimation errors were $(0.762\%,-1.939\%)$ of the exact $(\nu_1,\nu_2)=(1,0.6)$.

The results show that our method yielded a reasonable degree of accuracy even when the explicit PDF formulation is not available.

Given the above, it is clearly demonstrated that our proposed method has an advantage of not requiring any explicit PDF expression to estimate parameters. Although this advantage will become distinguished in nonlinear problems, we have restricted ourselves in this study to linear problems to demonstrate primal examples of our method.

\section{Conclusion}

In this study, we developed a new method that estimates the unknown parameter values of SDE models from PDF data without any explicit PDF formulation.
For this purpose, we used FPE residuals and developed a PDF fitness measure that tend to zero as the parameter values tend to exact values. Using the proposed measure, we estimated unknown parameter values of linear random vibration systems and obtained the following results.

As for a random linear vibration system with an additive noise only:
\begin{itemize}
  \item The independent parameters (i.e, the damping and stiffness) were estimated with estimation errors of under $1.7\%$.
  \item The relationship between the dependent parameters (i.e, the damping and noise strength) were extracted with estimation errors of under $2.9\%$.
\end{itemize}

As for a random linear vibration system with both additive and multiplicative noises:
\begin{itemize}
  \item The independent parameters (i.e, the additive and multiplicative noise-strengths) were estimated with estimation errors of under $2.0\%$.
\end{itemize}

The above results lead to the conclusion that our method achieved sufficient practical accuracy of the unknown parameter estimation even when the explicit PDF formulation is not available.
In future work, we plan to apply our method to nonlinear problems for which the advantage of our method will become distinguished.

\section*{Acknowledgment}

We would like to thank Prof. Hiroya Nakao for suggesting the topic treated in this paper.
This work was funded by JSPS KAKENHI Grant Numbers JP18H01391 and JP17H06552.


\end{document}